\documentstyle[prl,aps]{revtex}

\begin{document}
\input epsf
\draft
\twocolumn[\hsize\textwidth\columnwidth\hsize\csname@twocolumnfalse%
\endcsname

\title{Detecting Electronic States at Stacking Faults 
in Magnetic Thin Films by Tunneling Spectroscopy}

\author{A.L. V\'azquez de Parga$^{a*}$, F.J. Garc\'{\i}a-Vidal$^{b}$
and R. Miranda$^a$}

\address{$^a$Departamento de F\'{\i}sica de la Materia Condensada and
Instituto de Ciencia de Materiales ''Nicol\'as Cabrera"\\
Universidad Aut\'onoma de Madrid. 28049 Madrid. Spain.\\
$^b$Departamento de F\'{\i}sica Te\'orica de la Materia Condensada
\\ Universidad Aut\'onoma de Madrid. 28049 Madrid. Spain.}

\date{\today}
\maketitle

\begin{abstract}
Co islands grown on Cu(111) with a stacking fault at the interface
present a conductance in the empty electronic states larger than
the Co islands that follow the stacking sequence of the Cu
substrate. Electrons can be more easily injected into these
faulted interfaces, providing a way to enhance transmission in
future spintronic devices. The electronic states associated to the
stacking fault are visualized by tunneling spectroscopy and its
origin is identified by band structure calculations.
\end{abstract}

]


The new field of spin electronics or ''spintronics" refers to
electronic devices in which operation the electron spin is just as
important as its charge. For example, the two-component intrinsic
nature of spintronic devices might allow quantum computing in the
solid state. Recently an increasing number of devices that exploit
spin-polarized transport  (reading heads, sensors, magnetic random
access memories, metal base or semiconductor field-effect
transistors) have been proposed or built \cite{Prinz}. Some of
them take advantage of physical effects, such as the Giant
Magnetoresistence  (GMR) \cite{Baibich} or the Tunneling
Magnetoresistence (TMR) \cite{Modera}, which depend on the
spin-dependent flow of electrons with energies at the Fermi level.
Others, such as the spin-valve transistor (SVT) \cite{Monsma1},
rely on the ballistic transport of hot electrons with energies
0.5-2 eV above the Fermi level.

All these devices contain ferromagnetic/non ferromagnetic
interfaces. For instance, the base region of the first spin
transistors is formed by a Co/Cu multilayer
\cite{Monsma1,Monsma2}. The efficient transport of electrons
across these Co/Cu interfaces is crucial to the behavior of the
device. In fact, a severe drawback from the reported spin
transistors is the extremely low transmission of electrons with
0.5-2 eV above the Fermi level, that result in collector currents
ranging from 0.05$\%$ \cite{Monsma1} to 0.004$\%$ \cite{Monsma2}
of the emitter current.

We report in this Letter an experimental method to characterize
locally the transport of electrons at Co/Cu (111) interfaces based
on Scanning Tunneling Spectroscopy (STS). The method allows us to
identify the electronic states related to certain structural
arrangement, namely a stacking fault at the Co/Cu interface, that
enhance the transmission of electrons in the range 0.5-1.5 eV
above the Fermi level. The origin of the electronic states is
traced with the help of the calculated band structure. These
results suggest a way to manipulate the structure of these layered
structures that might result in further controlled enhancement of
the hot electron transport via properly tuned electronic states.

The experiments were carried out in a UHV chamber with a home made
STM, a rear-view LEED optics and Ultraviolet Photoelectron
Spectroscopy (UPS). The Cu(111) substrate was electrochemically
polished, sputter-annealed (500 eV, 800 K) for 5 minutes and
cooled to 300 K. The STM images (Fig. 1(a)) show straight
monoatomic steps and no traces of contaminants. The Co film was
deposited at room temperature from an electron bombardment source
with an evaporation rate of 0.4 ML per minute. The STM tips were
made of polycrystalline W wire electrochemically etched. In order
to perform the spectroscopic measurements the tip was cleaned in
situ by field emission as described elsewhere \cite{Flores}.

Figure 1(b) shows a topographic z(x,y) image of the surface after
evaporation of 0.4 ML of Co at room temperature on Cu(111). In
agreement with previous reports \cite{Figuera}, there are islands
of bilayer (4.2 $\rm{\AA}$) height over the Cu(111) surface, which
contain the deposited Co. Recently it has been proposed that the
islands are, in fact, three Cobalt layers deep, with the third
layer embedded within the Cu surface layer \cite{Pedersen}, as
schematically depicted in Fig. 1(c). The islands have triangular
shape with two different orientations, but, otherwise they appear
identical in topographic images. It was suggested \cite{Figuera}
that the shape and orientation of the Co islands could be
rationalized by considering that they nucleate respectively on the
two different threefold absorption sites of the surface as shown
in the Fig. 1(c) and that one type of compact step from the two
possible on Cu(111) surfaces, is kinetically favored. This would
imply that one of the orientations contains a stacking fault at
the interface. Although at this low coverage, diffraction
techniques lack enough spatial sensitivity to decide it, a recent
quantitative LEED study for a complete bilayer of Co/Cu(111)
\cite{Rath} concludes that 15$\%$ of the Co layer presents a
stacking-fault at the interface, while the rest of the bilayer
follows the FCC stacking sequence of the substrate. Accordingly,
the minority islands (pointing to the right in the Figure 1b)
would form a twin FCC structure and have a stacking-fault (faulted
islands), while the majority islands (pointing to the left) would
simply continue the FCC arrangement of the Cu substrate (unfaulted
islands).

The lower panel of Fig. 2 shows two differential tunneling
conductance images at different bias voltage, sampling the
occupied (Fig. 2(b)) and empty electronic states of the sample
(Fig 2(c)), respectively. The spectroscopic images have been
measured simultaneously with the constant-current topograph shown
in Fig. 2(a). From the images is clear that the tunneling
conductance for both type of islands is the same at negative
sample bias voltage, while it is clearly different at positive
sample bias voltage, with the faulted islands pointing to the
right showing larger conductance.

I-V curves were recorded on top of each type of islands to get a
more detailed information about such a prominent spectroscopic
feature. The experimental procedure was as follows. First an I-V
curve was measured on the substrate to verify that the tip was
clean, stable and reproduced the spectrum of the well known
surface state of Cu(111) \cite{ss}. Then we recorded I-V curves on
top of one type of Co island, checked that the tip did not change
by recording another spectrum on clean Cu(111) (between the
islands) and afterwards we measured on the other type of Co
island. In this way, differences in the spectra can be assigned to
changes in the electronic structure of the islands and not to
uncontrolled changes in the tip structure that often produce
localized electronic states in the tip apex which show up in the
I-V curves \cite{Flores}. Figure 3(a) shows the characteristic
surface state peak of clean Cu(111) at -0.4 V on the clean patches
of the surface and a rather featureless tunneling spectra on both
types of Co islands. The spectra on the islands are identical for
negative sample bias (occupied states) but different for positive
sample bias (empty states). In particular the minority (faulted)
islands present higher conductance for positive bias voltage in
agreement with Fig. 2.

In order to verify if a fault in the stacking sequence can provoke
such a dramatic change in the tunneling conductance of the
islands, band structure calculations were carried out using a
simple tight-binding formalism. In our theoretical framework,
islands are modeled by a laterally-infinite triple overlayer of Co
joined to a 20-layer Cu(111) film. This is reasonable since the
lateral size of the islands is large (100 $\rm{\AA}$) and the
spectra are taken in the center of the islands. Hopping
tight-binding elements for the Cu-Cu and Co-Co bonds are taken
from Ref. \cite{Papa} whereas we approximate the hopping Cu-Co by
the average of Cu-Cu and Co-Co interactions \cite{Vega}. For the
Co-Co interactions we have used the spin-polarized parameters. We
have not included atomic rearrangement in our calculations: this
effect is known to be small in metallic systems where atomic
density and local environment are preserved in the presence of a
stacking-fault. We have calculated the band structure of the slab
for two different kinds of stacking sequence of the cobalt
overlayer (ABCabc and ABCacb). As the energy levels and hopping
elements are exactly the same, {\bf the structural arrangement} is
therefore the only difference between the two cases.

The results of our calculation are shown in Fig. 3(b) and 3(c),
where the band structure of the slab is projected on the surface
Brillouin Zone and displayed along the
$\overline{\rm{\Gamma}}-\overline{\rm{M}}$ direction for the two
kind of stacking sequences: unfaulted in (b) and faulted with the
fault located between the first and second Co layers in (c). In
the $\overline{\rm{\Gamma}}-\overline{\rm{K}}$ direction the
results of our calculation are similar (not shown). First it is
interesting to point out that states that change depending on the
stacking sequence appear within the Cu(111) gap region
\cite{Yndurain}. Above the Fermi level (located at E=0 in the
figures), the band structure of the slab is governed by six bands
that have mainly $d_{xz}$ and $d_{yz}$ character ($z$ is normal to
the surface). The bands 5 and 6 have very low amplitude at the
surface. Bands 1 to 4, on the contrary, have strong weight at the
surface and are sensitive to the presence of a stacking-fault in
the Co-trilayer. The stacking fault breaks the symmetry in the
xy-plane and the resulting different hybridization between the
$d_{xz}-d_{yz}$ orbitals of the Co layer (that are effectively
decoupled of the Cu(111) bulk states) changes the dispersion
relation of these bands and provokes the splitting of them at the
$\rm{\Gamma}$ point as seen in Fig. 3. This distinct dispersion
with parallel momentum is a signal of a different spatial
distribution of the electronic states associated with these bands
in the two island orientations. For the faulted islands the first
four bands are remarkably flatter which implies that the
corresponding wavefunctions are more delocalized in the
z-direction (and then extending more into vacuum) that their
counterparts in the unfaulted islands. As a way of example, the
Local Density of States (LDOS) at 5$\rm{\AA}$ above the surface
associated with bands 3-4 at $\rm{\Gamma}$ point is 1.5 times
greater in the faulted islands than in the unfaulted ones. Then,
it is expected to obtain larger tunneling probabilities and hence
larger conductance when the tip is placed above a faulted island
that when it is above an unfaulted one. And this, over a large
energy range due to the splitting of the bands.

These results indicate that the structural arrangement right at
Co/Cu(111) interfaces determines the electronic states and the
conductance across such interfaces. This opens the possibility to
enhance hot-electron transport across Co/Cu multilayers by
properly engineering the geometric and electronic structure of the
Co layers.

In order to explore this possibility, we take advantage of
features that develop spontaneously during the growth of Co on
Cu(111). Although the majority islands of Co continue the FCC
stacking sequence of the Cu substrate upon further Co deposition,
the film quickly starts to adopt the HCP, bulk-like structure
\cite{Camarero1}. In effect, Co atoms in the third layer on choose
between the two threefold adsorption sites with almost equal
probability giving rise to a pyramidal-shaped morphology
\cite{Figuera}, with pyramids presenting the FCC or HCP stacking
sequence as deduced from surface X-ray diffraction data
\cite{Camarero1}. We show below that their different electronic
structure can be locally identified by STS.

For tips showing a strong Cu(111) surface state band-edge, only
two predominant types of spectra were observed at different points
on 6 ML thick Co films. These are shown in Fig. 4(a) and (b).
Similarly to the initial islands, the conductance corresponding to
the occupied states is almost identical, but clear differences are
seen in the empty states. In panels (c) and (d) of the same
figure, we present the calculated total LDOS around
$\overline{\rm{\Gamma}}$ for a Cobalt film 6 ML thick with FCC and
HCP stacking sequences, respectively, at different distances from
the surface. By simply comparing Figs. 4c-d with Figs. 4a-b, we
can conclude that pyramids showing spectra like (a) follow a FCC
stacking whereas spectra like (b) are linked to HCP stacking.
Noting that the FCC phase shows a larger LDOS than the HCP phase,
we could suggest that varying growth conditions (e.g. the use of
surfactants, coevaporation or pulsed laser deposition) that could
stabilized the FCC rather than the HCP phase of Co might yield new
ways to enhance the transmission in Co/Cu multilayers.

The physical origin of the differences between the electronic
structure of both stacking sequences is the same than for the
thinner islands. When the stacking sequence is FCC, bands
associated to d$_{xz}$-d$_{yz}$ orbitals of the Co layer are
degenerated at the $\Gamma$ point in a way very much similar to
the case of unfaulted islands (see Fig. 3b). Hence a peak in the
LDOS at 0.5 eV above Fermi level is expected (see Fig. 4c).
However when the stacking sequence is HCP, the different
hybridization of those orbitals provokes the opening of a gap at
the $\Gamma$ point also in accordance of the case of faulted
islands (see Fig. 3c). Therefore a LDOS with two peaks located
below and above 0.5 eV is obtained as seen in Fig. 4d. The
agreement between the experimental data and the calculated density
of states is remarkable. Even the relative intensity between the
FCC and HCP films in the calculations and experiments is
conserved.

In conclusion, we have found that the differential tunneling
conductance is higher on Cobalt islands grown on Cu(111) that have
a stacking fault at the interface than on islands that continue
the FCC-stacking of the substrate. Using a simple theoretical
model, in which the stacking sequence is the only difference
between both type of islands, we have identified the electronic
states responsible for the enhanced conductance. These results
open a possibility to enhanced the transmission of hot electrons
across magnetic/non magnetic interfaces by properly tuning the
geometric and electronic structure of these artificial materials.

This work has been financed by grants MAT98-C04-02
and PB97-0031.

\begin{figure}
\caption[]{(a) (1000$\times$1000$\rm{\AA}$) STM image of
the clean Cu(111) surface. (b)
(1500$\times$1500$\rm{\AA}$) 0.6 ML of Cobalt deposited at room
temperature with an evaporation rate of 0.4 ML per minute. Two
types of bilayer-high (4.2 $\rm{\AA}$) triangular islands appear on the surface.
 The islands are marked with "{\bf u}" (unfaulted) and "{\bf f}" (faulted)
depending on the stacking sequence. (c) Model proposed for the two
types of islands shown in panel (b). The islands are three layers
high with one layer within the copper surface. The unfaulted ("u")
islands follow the FCC stacking sequence and the faulted ones
("f") present an stacking-fault.} \label{fig1}
\end{figure}

\begin{figure}
\caption[]{(a) (500$\times$500$\rm{\AA}$) STM image after
deposition of 0.4 ML of Cobalt on Cu(111) at room temperature.
There is no difference in the height of both type of islands in
the topographic images within the noise. Panels (b) and (c) shown
spectroscopic images, (dI/dV)(x,y), of the region shown in (a),
measured with the feedback loop stabilized at a sample bias of (b)
-1.5 V and (c) +1.5 V. The bias was modulated (70 meV) with a
frequency (8 kHz) higher than the feedback loop response and the
ac-component of the current was measured by a lock-in detector.
Bright areas correspond to high differential conductance.}
\label{fig2}
\end{figure}

\begin{figure}
\caption[]{(a) Experimental differential conductance for a clean
area of the Cu(111) surface (solid line),for a Co island that
follows the Cu stacking (unfaulted) (dashed line) and for a Co
island with a stacking fault (faulted) (dotted line). Energy bands
in the [111] direction for (b) the unfaulted Co island and (c) for
the island that presents a stacking-fault at the interface.}
\label{fig3}
\end{figure}

\begin{figure}
\caption[] {Panels (a) and (b) present tunneling spectra measured
over different places in a 6 ML-thick film of Co grown on Cu(111)
at room temperature. Panels (c) and (d) display the calculated
LDOS at different distances from the surface, integrated over
k$_{\|}$ in a circular region  around $\overline{\Gamma}$, out to
$\frac{1}{4}$ of the $\overline{\Gamma}$-$\overline{\rm{M}}$
distance, for 6 ML of Co following (c) an FCC stacking sequence
and (d) an HCP stacking one. The prominent peaks at $\sim$ 0.5 eV
correspond to unoccupied minority electron states. The calculated
LDOS has been smoothed with a Gaussian of 0.2 eV FWHM to simulate
experimental resolution.} \label{fig4}
\end{figure}

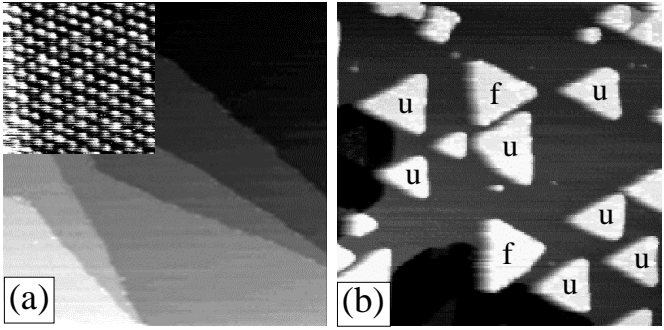
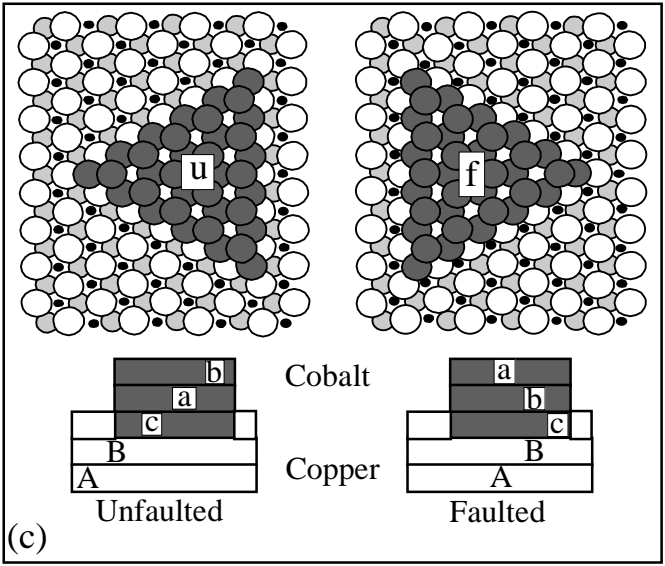
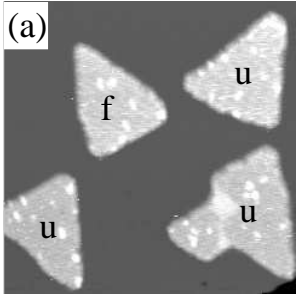
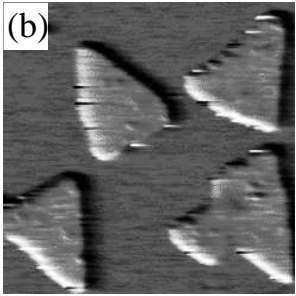
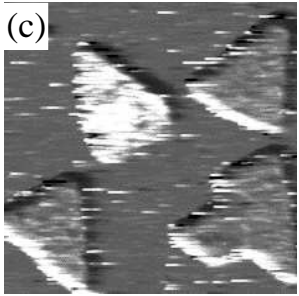
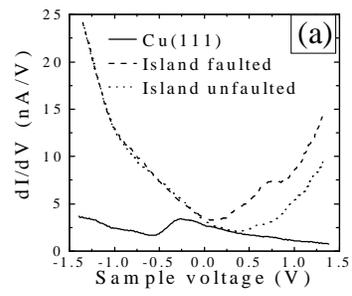
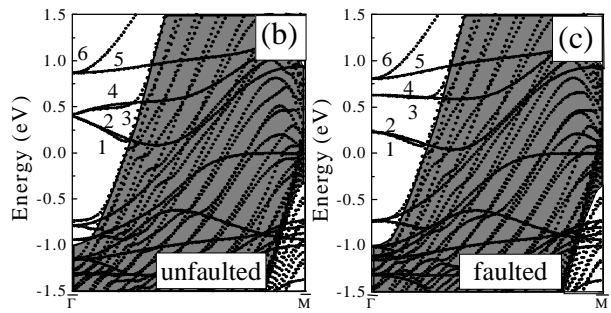
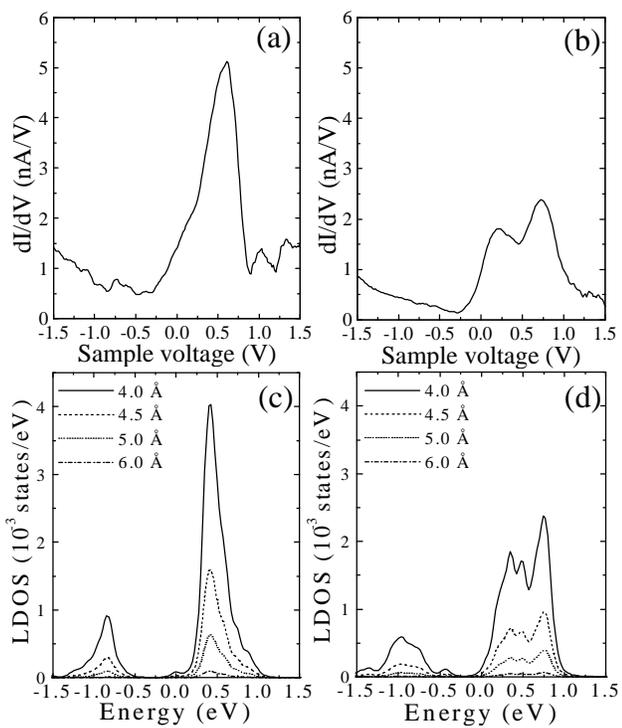
\end{document}